\documentclass[fleqn,twoside]{article}
\usepackage{espcrc2}
\usepackage{epsfig}
\usepackage{amssymb}
\usepackage[figuresright]{rotating}

\newcommand{\AmS}{{\protect\the\textfont2
  A\kern-.1667em\lower.5ex\hbox{M}\kern-.125emS}}

\newcommand{\ksea}{\mbox{$\kappa^{\rm sea}$}}
\newcommand{\kval}{\mbox{$\kappa^{\rm val}$}}

\newcommand{\ltap}{\;\raisebox{-.5ex}{\rlap{$\sim$}} \raisebox{.5ex}{$<$}\;}
\newcommand{\gtap}{\;\raisebox{-.5ex}{\rlap{$\sim$}} \raisebox{.5ex}{$>$}\;}

\newcommand{\be}{\begin{equation}}
\newcommand{\ee}{\end{equation}}
\newcommand{\bea}{\begin{eqnarray}}
\newcommand{\eea}{\end{eqnarray}}

\newcommand{\figurebox}[2]{\fbox{\vbox to#2in{\hbox to #1in{\hfil}\vfil}}}
\newcommand{\err}[2]{${\scriptstyle {}^{+{#1}}_{-{#2}}}$}

\newcommand{\bm}[1]{\mbox{\boldmath ${#1}$}}

\newcommand{\rvec}{\bm{r}}

\def\lsi{\raise0.3ex\hbox{$<$\kern-0.75em\raise-1.1ex\hbox{$\sim$}}}
\def\gsi{\raise0.3ex\hbox{$>$\kern-0.75em\raise-1.1ex\hbox{$\sim$}}}

\title{Effects of Dynamical Quarks in UKQCD Simulations}

\author{Chris Allton\\
\vspace{3mm}
{\em UKQCD Collaboration}\\
\vspace{3mm}
Department of Physics, University of Wales Swansea,
Singleton Park, Swansea SA2 8PP, United Kingdom}

\begin{document}

\begin{abstract}

Recent results from the UKQCD Collaboration's dynamical simulations
are presented. The main feature of these ensembles is that they
have a fixed lattice spacing and volume, but varying sea quark
mass from infinite (corresponding to the quenched simulation)
down to roughly that of the strange quark mass.
The main aim of this work is to uncover dynamical quark effects
from these ``matched'' ensembles.
We obtain some evidence of dynamical quark effects in the static
quark potential with less effects in the hadronic spectrum.

\end{abstract}

\maketitle


\section{Introduction}

The UKQCD Collaboration has embarked upon a study of unquenching
effects in QCD lattice simulations particularly in the static
quark potential, light hadron spectrum and topological sector.
The philosophy that UKQCD uses in this study is to simulate
at a variety of sea quark masses, $m^{sea}$, but with a fixed
the lattice spacing, $a$. This is in order to maintain,
as far as possible, constant lattice systematic effects due
to the finite-ness of $a$ and the volume.

It is well known that the lattice cut-off, $a$, is a function
of {\em both} the gauge coupling, $\beta$, and the dynamical
quark mass, $m^{sea}$ (see e.g. \cite{csw176}).
For this reason, the philosophy we have chosen is to simulate
at points in the $(\beta,m^{sea})$ plane which have constant $a$.
We term this the ``matched'' trajectory.
The hope is that this procedure will disentangle lattice systematic
effects (due to the finite-ness of $a$ and volume) with
unquenching effects.
Any variation of a physical quantity along this matched
trajectory can be attributed to unquenching effects
rather than, e.g. ${\cal O}(a^2)$ effects.

This paper presents an overview of this work. See \cite{csw202}
for a full description.


\section{Simulation Details}

The standard Wilson gauge action was used together with the
Clover ${\cal O}(a)$-improved fermion action.
The coefficient, $c_{SW}$, used in the Clover term was non-perturbatively
determined by the Alpha Collaboration \cite{alpha}.

A range of $(\beta,\ksea)$ values were chosen in order to maintain
a constant value of the lattice spacing. This used the technology
of \cite{aci}.
The parameter values for the simulations are displayed in
Table \ref{tb:params}. The lattice volume used was $16^3\times32$ throughout.

Table \ref{tb:params} lists also the lattice spacing obtained
from the Sommer scale, $r_0$ \cite{sommer}.
The first error listed is statistical, and the second (shown as \err{x}{y})
is the systematic error from variations in the fit used for the
static quark potential. The central values quoted were obtained
using all potential data satisfying $\sqrt{2}\leq \rvec\leq 8$.
It can be seen that the last four simulations
listed (i.e. those at $\ksea=0.1350, 0.1345, 0.1340$ and $0$)
are matched to within errors, with the $\ksea=0$ simulation
corresponding to a quenched run.

The simulations at $\ksea=0.1355$ and $0.13565$ were performed in order
to study lighter sea quark masses and do not lie on the above matched
trajectory.

Full details of autocorrelation times and other algorithmic issues
can be found in \cite{csw202}.


\begin{table*}[tb]
\caption{Lattice parameters together with measurements of $a$ and quark masses.}
\label{tb:params}
\begin{center}

\begin{tabular}{lllllll}
$\beta$ & $\ksea$ & $c_{SW}$ & \#conf. &  
\multicolumn{1}{c}{ $a_{r_0}$ [Fm]} &
\multicolumn{1}{c}{ $a_{J}$ [Fm]} &
\multicolumn{1}{c}{ $M_{PS}^{unitary}/M_V^{unitary}$}
 \\
\hline
5.20& 0.13565&2.0171 & 244 & 0.0941(8)\err{13}{0}  & -                & - \\
5.20& 0.1355 &2.0171 & 208 & 0.0972(8)\err{7}{0}   & 0.110\err{ 4}{ 3}& 0.578\err{13}{19}\\
\hline
5.20& 0.1350 &2.0171 & 150 & 0.1031(09)\err{20}{1} & 0.115\err{ 3}{ 3}& 0.700\err{12}{10}\\
5.26& 0.1345 &1.9497 & 101 & 0.1041(12)\err{11}{10}& 0.118\err{ 2}{ 2}& 0.783\err{ 5}{ 5}\\
5.29& 0.1340 &1.9192 & 101 & 0.1018(10)\err{20}{7} & 0.116\err{ 3}{ 4}& 0.835\err{ 7}{ 7}\\
\hline 
5.93& 0      &1.82   & 623 & 0.1040(03)\err{4}{0}  & 0.1186\err{17}{15}& 1   \\
\end{tabular}
\end{center}
\end{table*}


\section{The static potential}
\label{sec:pot}

The static quark potential was determined for all our ensembles
using the method originally proposed in \cite{cmi} and using
a variational basis of ``fuzzed'' links \cite{ape}.


\begin{figure}[htp]
\begin{center}
\epsfig{file=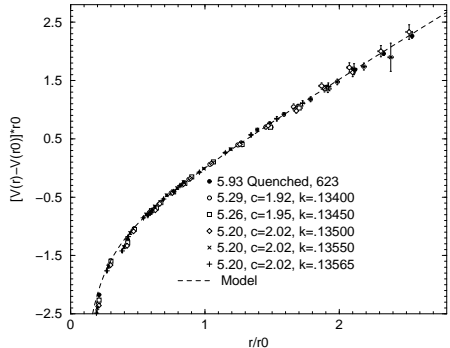,height=5.5cm,angle=0}
\end{center}
\label{fig:r0V}
\vspace{-10mm}
\caption{The static QCD potential expressed in units of $r_0$.
The dashed curve is a string model described in the text.
}
\end{figure}


Figure 1 plots the static potential in 
units of $r_0$. 
The zero of the potential has been set at $r=r_0$.
The data are well described by the
universal bosonic string model potential~\cite{Luscher} which
predicts
\begin{eqnarray}\nonumber
  \big[V(r)-V(r_0)\big]r_0 &=& (1.65-e)\left(\frac{r}{r_0}-1\right)\\
  &-&e\left(\frac{r_0}{r}-1\right)\, .
\label{eq:potstring}
\end{eqnarray}
Of course, the fact that the scaled potential measurements all have the same
value and slope at $r=r_0$ simply reflects the definition of $r_0$.
Figure 2 shows the deviations from this
model potential for the matched ensembles. Here $e=\pi/12$~\cite{Luscher}. 

Overall there is no discernable difference between the ensembles at
distance $r\approx r_0$. Furthermore, the data follow the string
model very well. However, at shorter distances, $r<0.5r_0$,
there is a deviation from the string model, and a variation amongst
the ensembles. This seems to be systematic in the sea quark mass --
the deviation from the string model increases as $m^{sea}$ decreases.
In fact, careful correlated fits of the potential show that the parameter
$e$ in Eq.(\ref{eq:potstring}) increases by
18\err{13}{10}\% in going from the quenched to $\ksea=.13500$ data.
We note that the ensembles in Figure 2 are matched,
and therefore we can exclude lattice systematics from the effects
that we have observed.
(Similar findings in the case of two flavours of Wilson fermions have
been reported by the SESAM-T$\chi$L collaboration~\cite{txl} where
an increase of $16-33\%$ was found.)
Finally we note that there is no evidence of string breaking in
our static quark potential. However the distance scales covered
$r \ltap 1.3 fm$ is not large.


\begin{figure}[htp]
\begin{center}
\epsfig{file=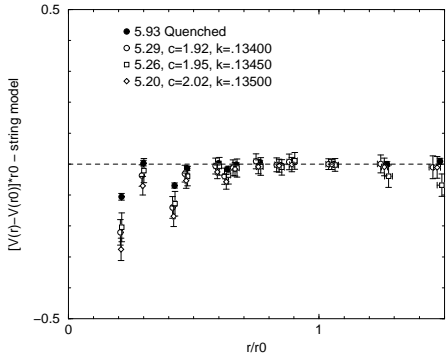,height=5.5cm,angle=0}
\end{center}
\label{fig:PotDiff}
\vspace{-10mm}
\caption{The difference between the static QCD potential expressed in
  physical units and the prediction of the string model described in
  the text. For clarity, only data from the matched ensembles are shown.}
\end{figure}



\section{Hadronic Spectrum}


In this section one of our main aims will be to uncover
unquenching effects in the light hadron spectrum.
Because we have a {\em matched} data set, any variation
amongst our ensembles can be attributed to unquenching effects. However, 
the task of identifying variations is likely to be
hard for those quantities which are primarily sensitive
to physics at the same scale as that used to define the matching
trajectory in the $(\beta,\ksea)$ parameter space
($r_0$ in this case). This is expected to be the
case for the hadron spectrum considered here where the
quark masses are still relatively heavy.

Two-point hadronic correlation functions were produced for each of the
ensembles using interpolating operators for pseudoscalar, vector,
nucleon and delta channels as described in~\cite{ukqcd1}.
Mesonic correlators were constructed using both degenerate and
non-degenerate valence quarks, whereas only degenerate valence quarks
were used for the baryonic correlators.

In the following, we review the main fitting procedures 
which were used to obtain the light hadron spectrum results.
Further details of the fitting procedure can be found in
\cite{csw202}.

We used the fuzzing procedures of~\cite{Lacock} to generate
correlators of the type LL, FL and FF where F denotes fuzzed, and L
local operators.

Correlated fits were used throughout the fitting analysis of the
correlation functions with the eigenvalue smoothing technique of
of~\cite{cmi2} employed.
A {\em factorising fit} was performed which combined the
three fuzzed correlators LL, FL and FF together to extract
a better estimate of the ground state parameters.

Full details of the fitting procedure can be found in \cite{csw202}.



\subsection{The J parameter}
\label{sec:J}

In Figures 3 and 4 the vector meson
masses and hyperfine splittings are plotted against
the corresponding pseudoscalar masses for all the datasets.  It is
difficult to identify an unquenching signal from these plots - the data
seem to overlay each other.  Note that in \cite{csw176}, it was
reported that there was a tendency for the vector mass to {\em
 increase} as the sea quark mass {\em decreases} (for fixed
pseudoscalar mass). The observations for the present {\em matched}
dataset imply that this may have been due to either an ${\cal O}(a)$
effect (since the dataset in \cite{csw176} was not fully
improved at this level) or a finite volume effect.  The conclusion
therefore is that it is important to run at a fixed $a$ in order to
disentangle unquenching effects from lattice artifacts or finite
volume effects.


\begin{figure}[htp]
\begin{center}
\epsfig{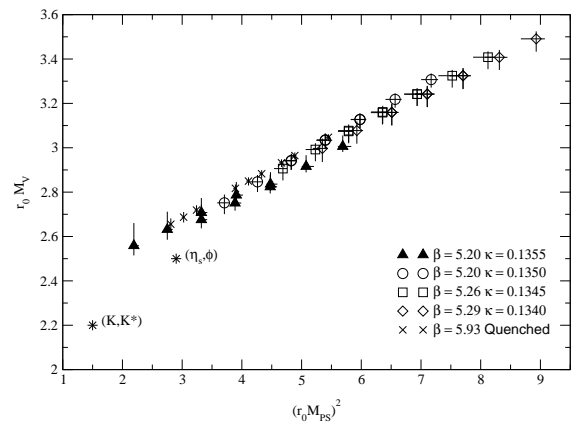}
\end{center}
\label{fig:mesons}
\vspace{-10mm}
\caption{
Vector mass plotted against pseudoscalar mass squared in units of $r_0$,
together with the experimental data points.
}
\end{figure}


A possible explanation for the lack of convincing unquenching effects in
our meson spectrum is the following. Our matched ensembles are defined
to have a common $r_0$ value, so any physical quantity that is
sensitive to this distance scale (and the static quark potential
itself) will also, by definition, be matched. Our mesons, because they
are composed of relatively heavy quarks, are examples of such
quantities, and this is a possible reason why there is no significant
evidence of unquenching effects in the meson spectrum.

A further point regarding hyperfine splitting in Figure 4
is that the lattice data for the {\em matched} ensembles tends to
flatten as the sea quark mass decreases. (The quenched data has a distinctly
negative slope, whereas the $\ksea=0.1350$ data is flat.) Thus the
lattice data is tending towards the same behaviour as the experimental
data which lies on a line with {\em positive} slope (independent of
the value used for $r_0$).
This behaviour is apparently spoiled by the unmatched run with $\ksea=0.1355$
(see Figure 4) which has a clear {\em negative} slope.
However, the $\ksea=0.1355$ data does not satisfy the finite volume bound
of \cite{csw176}, and therefore we can attribute the trend in this
data to a lattice artefact.

We now study the $J$-parameter defined as \cite{Lacock:1995tq}
\be
J = M_V \frac{dM_V}{dM_{PS}^2} \bigg|_{K,K^\ast}.
\label{eq:J}
\ee


\begin{figure}[htp]
\begin{center}
\epsfig{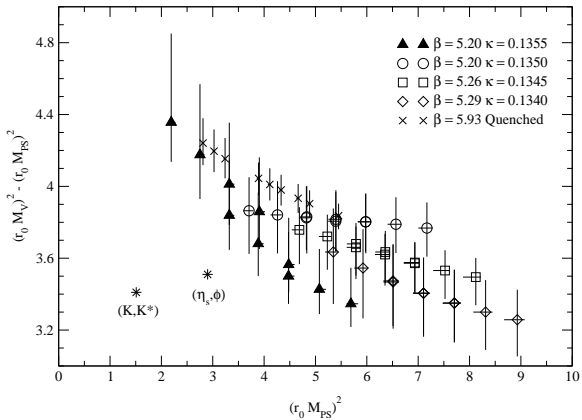}
\end{center}
\label{fig:hyperfine}
\vspace{-10mm}
\caption{The hyperfine splitting.}
\end{figure}


In the context of dynamical fermion simulations, this parameter can be
calculated in two ways. The first is to define a partially quenched
$J$ for each value of the sea quark mass. In this case, the derivative
in (\ref{eq:J}) is with respect to variations in the valence quark
mass (with the sea quark mass fixed). The second approach is to define
$J$ along what we will term the `unitary' trajectory,
i.e. along $\ksea = \kval$.
In Table \ref{tb:J}, the results from both methods are given.
These values of $J$ are around 25\% lower than
the experimental value $J_{expt} = 0.48(2)$.


\begin{table}[*htbp]
\begin{center}
 \begin{tabular}{ccccc}
  & $\beta$ & $\ksea$ & $J$ & \\
\hline
\multicolumn{5}{c}{\bf First Approach} \\
  & 5.2000 & 0.1355 & 0.32\err{ 2}{ 4} & \\
\hline
  & 5.2000 & 0.1350 & 0.393\err{10}{ 9} & \\
  & 5.2600 & 0.1345 & 0.365\err{ 6}{ 6} & \\
  & 5.2900 & 0.1340 & 0.349\err{ 7}{ 8} & \\
\hline
  & 5.9300 & 0.0000 & 0.376\err{ 9}{12} & \\
\hline
\multicolumn{5}{c}{\bf Second Approach} \\
  & - & - & 0.35\err{ 2}{ 2} & \\
\hline
\multicolumn{5}{c}{\bf Third Approach} \\
  & - & - & 0.43\err{ 2}{ 2} & \\
\end{tabular}
\end{center}
\caption{$J$ values from the various approaches as described in the text.}
\protect\label{tb:J}
\end{table}


Finally we note that the physical value of $J$ (i.e. that which
most closely follows the procedure used to determine the
experimental value of $J_{expt} = 0.48(2)$),
should be obtained from extrapolating the results from the first approach
to the physical sea quark masses. We call this the third approach.
In order to perform this extrapolation, we extrapolate the three matched
dynamical $J$ values
obtained from the first approach linearly in $(M_{PS}^{\rm unitary})^2$
to $(M_{PS}^{\rm unitary})^2=0$.
$M_{PS}^{\rm unitary}$ is the pseudoscalar meson mass at the unitary
point.
The value for $J$ from the third approach is presented in Table \ref{tb:J}
and we note that it is approaching the experimental value for $J$.

The results from all three approaches are plotted
in Figure 5, together with the experimental result.
There is some promising evidence that the lattice estimate of $J$ increases
towards the experimental point as the sea quark mass decreases (see
the $J$ value from approaches 1 and 3).


\begin{figure}[htp]
\begin{center}
\epsfig{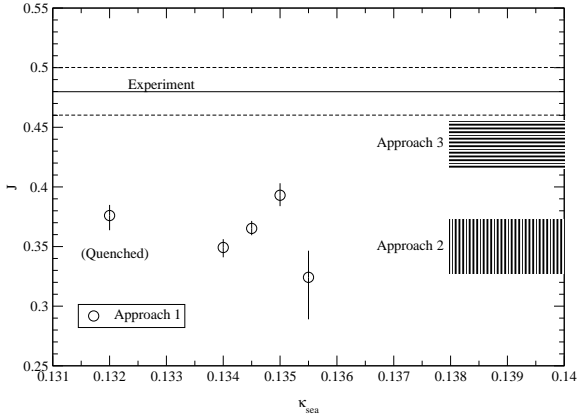}
\end{center}
\label{fig:J}
\vspace{-10mm}
\caption{
$J$ versus $\kappa^{sea}$ using the approaches as described in the text.
Note that the quenched data points have been plotted at $\ksea=0.132$
for convenience.
Approaches 2 \& 3 are obtained after a chiral extrapolation and are
shown as banded regions.
The experimental value $J = 0.48(2)$ is also shown.
}
\end{figure}


Recently there has been a proposed ansatz for the functional form of
$M_V$ as a function of $M_{PS}^2$ \cite{Leinweber:2001ac}.  Although
this ansatz would be interesting to pursue, all our data have
$M_{PS}/M_V \gtap 0.6$, and for this region, the ansatz of
\cite{Leinweber:2001ac} is linear to high precision.  Therefore
we have chosen to interpolate our data with a simple linear function and
await more chiral data before using the ansatz of
\cite{Leinweber:2001ac}.



\subsection{Lattice spacing}
\label{sec:lspac}
This subsection presents a determination of $a$ from
the meson spectrum which complements that from $r_0$.

A common method of determining $a$ from the meson spectrum uses the
$\rho$ mass. However, this requires the chiral extrapolation of the
vector meson mass down to (almost) the chiral limit which,
as was discussed in the previous subsection, may be problematic.
An alternative method of extracting the lattice spacing using the vector
meson mass at the {\em simulated} data points (i.e.  without any
chiral extrapolation) was given in \cite{Allton:1997yv}.  Using this
method, we obtain the lattice spacing values as shown in Table
\ref{tb:params} labelled $a_J$. Note that these are in general 10-15\% larger than the
values from Section \ref{sec:pot} where the lattice spacing was
determined from $r_0$. A possible explanation for this discrepancy is
that the potential and mesonic spectrum are contaminated with
different ${\cal O}(a^2)$ errors (or that the value $r_0 = 0.49$ fm
is 10-15\% too small!).

In order to investigate unquenching effects in the meson spectrum,
we define the quantity
\be
\delta_{i,j}(\beta,m^{sea}) = 
1 - \frac{a_i(\beta,m^{sea})}{a_j(\beta,m^{sea})},
\label{eq:delta_ij}
\ee
where $a_i$ is the lattice spacing determined from the physical quantity
$i = \{ M_\rho, M_K, f_\pi \ldots \}$.
Note that when $\delta_{i,j}=0$, the lattice prediction of $M_i$ with scale
taken from $M_j$ agrees with experiment.
Thus $\delta$ is a good parameter to study unquenching effects.
We expect that $\delta_{i,j}(\beta,m^{sea}) = {\cal O}(a^2)$ since we
are working with a non-perturbatively improved clover action.

In Figure 6, $\delta_{i,j}$ is plotted against
$(aM_{PS}^{\rm unitary})^{-2}$ for the matched datasets.
In this plot we have fixed $j = r_0$ and
the various physical quantities $i$ are $\sqrt\sigma$ (the string tension)
and the hadronic mass pairs $(M_K*,M_K)$ \& $(M_\rho,M_\pi)$.
The method used to determine the scale $a_i$ from these mass
pairs is that of \cite{Allton:1997yv}.
It is worth noting that the experimental point on this same plot would occur
at an $x-$co-ordinate (depressingly) of
$(aM_{PS}^{\rm unitary})^{-2} = (aM_\pi)^{-2} \approx 200$.


\begin{figure}[htp]
\begin{center}
\epsfig{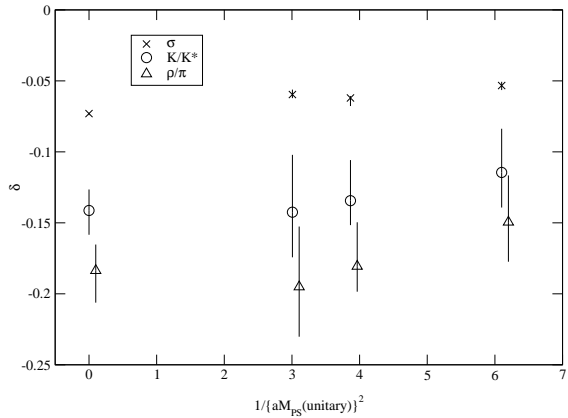}
\end{center}
\label{fig:delta}
\vspace{-10mm}
\caption{
$\delta_i$ as a function of $1/(aM^{unitary}_{PS})^2$ for $i =
\sqrt{\sigma}$ and the mass pairs $(M_K*,M_K)$ \& $(M_\rho,M_\pi)$.
$\delta_i$ is defined in eq.(\ref{eq:delta_ij}) with $j = r_0$.
}
\end{figure}


Figure 6 does not show signs
of unquenching for quantities involving the hadronic spectrum
i.e. the mass pairs $(M_K*,M_K)$ \& $(M_\rho,M_\pi)$.
However, there is evidence of unquenching effects when comparing the
scale from $r_0$ with that from $\sqrt\sigma$ since the quenched value
of $\delta_{\sqrt\sigma}$ is distinct from the dynamical values.

One may wonder if the the $\delta$ values may have been distorted by
not choosing the simulation parameters $(\beta,m^{sea})$
exactly on the matched trajectory.
In order to obtain a rough estimate of the effect of mis-matched value
of $\beta$, we use the
renormalisation group inspired ansatz for $a_i$ \cite{Allton:1996kr,Allton:1997dn}:
\begin{equation}
a_i(g_0^2) = \Lambda^{-1} f_{PT}(g_0^2)
\times  \left[ 1 + X_i f_{PT}(g_0^2)^{n_i} \right],
\label{eq:ldpt_fit}
\ee
where $f_{PT}(g^2)$ is the usual asymptotic scaling function obtained
from integrating the $\beta$-function of QCD and $X_i$ is the coefficient
of the ${\cal O}(a^n)$ lattice systematic.
The functional form for $a(g_0^2)$ was originally applied for the quenched
theory, but let us assume that it can also be applied in the unquenched case.
Using Eq.(\ref{eq:ldpt_fit}), we see that a mismatch in $\beta$ of
$\Delta\beta$ would lead to a relative error in $\delta$ of
\begin{equation}
\frac{\delta(\beta+\Delta\beta) - \delta(\beta)}{\delta(\beta)}
\approx -3 \Delta\beta.
\label{eq:deltabeta}
\end{equation}
This shows that even an error in $\beta$ of as much as
$\Delta\beta \approx 0.01$
introduces a relative error in $\delta(\beta)$ of only 3\%,
ruling out the possibility that a significant distortion in $\delta$
could have occurred due to a mismatching in $\beta$.



\subsection{Edinburgh Plot}
\label{sec:edinplot}

In Table \ref{tb:params} the ratios $M_{PS} / M_{V}$ are displayed
for the unitary case \ksea = \kval. As can be seen the simulated
data is a long way from the experimental value $M_\pi / M_\rho= 0.18$.
Figure 7 shows the \lq Edinburgh plot\rq{} ($M_N/M_V$ v.s.
$M_{PS}/M_V$) for all the data sets.  There is no significant
variation within the dynamical data as the sea quark mass is changed,
but the dynamical data does tend to lie above the (matched) quenched
data.  This latter feature may be indicative of finite volume effects
since these are expected to be larger in full QCD compared to the
quenched case \cite{Ukawa:1993hz}.


\begin{figure}[htp]
\begin{center}
\epsfig{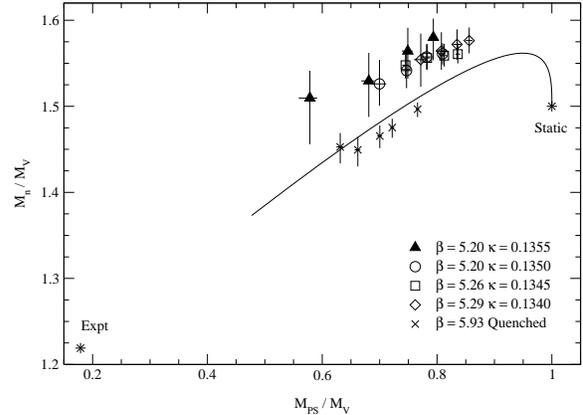}
\end{center}
\label{fig:edin}
\vspace{-10mm}
\caption{
The Edinburgh plot for all the data sets. All degenerate $\kval$
correlators have been included. The phenomenological curve (from
\protect\cite{Ono:1978}) 
has been included as a guide to the eye.
}
\end{figure}




\section{Chiral Extrapolations}

In \cite{csw202} we used 3 approaches to perform the chiral
extrapolations of hadron masses: 
``Pseudo-Quenched'';
``Unitary Trajectory'';
and a ``Combined Chiral Fit''.
In this publication we will refer only to the last approach.
We take
\[
\hat{M}(\ksea;\kval)
\]
\[
\;\;\;\;\;\;\;\;\;\;\;\;= A(\ksea) + B(\ksea) \hat{M}_{PS}(\ksea;\kval)^2
\]
\[
\;\;\;\;\;\;\;\;\;\;\;\;= A_0 + A_1 \hat{M}_{PS}(\ksea;\ksea)^{-2}
\]
\[
+ \left[ B_0 + B_1 \hat{M}_{PS}(\ksea;\ksea)^{-2} \right]
\hat{M}_{PS}(\ksea;\kval)^2,
\]
using the nomenclature $\hat{M} \equiv aM$.
The first argument of $M(\ksea;\kval)$ refers
to the sea quark and the second to the valence quark.

The results of these extrapolations are shown in
Table~\ref{tb:chiral}. We stress that this functional form for the
extrapolation is not motivated by theory, but is used as a numerical
analysis technique in order to test for evidence of
unquenching effects. As can be seen from Table~\ref{tb:chiral}, the
parameters $A_1$ and $B_1$ are compatible with zero (to $2\sigma$)
and therefore we conclude that there is no evidence of unquenching
effects (i.e. there is no statistical variation in the fit
parameters $A$ and $B$ with sea quark mass $\ksea$).


\begin{table}[*htbp]
\begin{center}
 \begin{tabular}{lcccc}
hadron  & $A_0$ & $A_1$ & $B_0$ & $B_1$ \\
\hline
Vector       & .492\err{10}{ 9} & -0.004\err{ 2}{ 3} & 0.61\err{ 4}{ 4} &  .015\err{ 9}{ 7} \\
Nucleon      & .663\err{13}{15} &  0.006\err{ 3}{ 4} & 1.23\err{ 6}{ 6} & -.001\err{1}{1} \\
Delta        & .84\err{ 2}{ 2}  & -0.002\err{ 5}{ 5} & 0.91\err{ 8}{ 9} &  .02\err{ 2}{ 2} \\
\end{tabular}
\end{center}
\caption{Fit parameters from the Chiral Extrapolations}
\label{tb:chiral}
\end{table}



\section{Conclusions}

This paper attempts to uncover unquenching effects in the dynamical
lattice QCD simulations at a fixed (matched) lattice spacing (and
volume) and various dynamical quark masses. This approach allows
a more controlled study of unquenching effects without the possible
entanglement of lattice and unquenching systematics.

We see some sign of unquenching effects in the static quark potential
at short distance, but no significant sign of unquenching effects in the
meson spectrum. This is presumably since our dynamical quarks are
relatively massive, and so the meson spectrum is dominated by the
static quark potential. This potential is, by definition, matched
amongst our ensembles at the hadronic length scale $r_0$, and so
any variation of the meson spectrum within our matched ensemble
must surely be a ``higher'' order unquenching effect which is beyond
our present statistics.

It is interesting to note that other work (using the same ensembles) has shown
interesting unquenching effects in the glueball and topological sector
\cite{csw202}.



\section{Acknowledgements}

The author wish to thank all of his collaborators in UKQCD.
The support of the Particle Physics and Astronomy Research
Council is gratefully acknowledged.




\begin{thebibliography}{99}

\bibitem{csw176} UKQCD Collaboration, C.R.Allton {\it et al}.,
Phys.Rev. {\bf D60} (1999) 034507,
{\tt hep-lat/9808016}.

\bibitem{csw202} UKQCD Collaboration, C.R.Allton {\it et al}.,
{\tt hep-lat/0107021}.

\bibitem{alpha}
ALPHA Collaboration, K.~Jansen and R.~Sommer,
Nucl. Phys. {\bf B530}, 185 (1998), {\tt hep-lat/9803017}.

\bibitem{aci}
UKQCD Collaboration, A.~C. Irving {\em et~al.},
Phys. Rev. {\bf D58}, 114504 (1998), {\tt hep-lat/9807015}.

\bibitem{sommer}
R.~Sommer, Nucl. Phys. {\bf B411}, 839 (1994), {\tt hep-lat/9310022}.

\bibitem{cmi}
C.~Michael, Nucl. Phys. {\bf B259}, 58 (1985);
S.~Perantonis, A.~Huntley, and C.~Michael,
Nucl. Phys. {\bf B326}, 544 (1989).

\bibitem{ape}
APE Collaboration, M.~Albanese {\em et~al.},
Phys. Lett. {\bf B192}, 163 (1987).

\bibitem{Luscher}
M.~Luscher,  Nucl. Phys. {\bf B180}, 317 (1981).

\bibitem{txl}
SESAM Collaboration, G.~S. Bali {\em et~al.},
Phys. Rev. {\bf D62}, 054503 (2000), {\tt hep-lat/0003012}.

\bibitem{ukqcd1}
UKQCD Collaboration, C.~R. Allton {\em et~al.},
Phys. Rev. {\bf D49}, 474 (1994), {\tt hep-lat/9309002}.

\bibitem{Lacock}
UKQCD Collaboration, P.~Lacock, A.~McKerrell, C.~Michael,
I.~M. Stopher, and P.~W. Stephenson,
Phys. Rev. {\bf D51}, 6403 (1995), {\tt hep-lat/9412079}.

\bibitem{cmi2}
C.~Michael and A.~McKerrell,
Phys. Rev. {\bf D51}, 3745 (1995), {\tt hep-lat/9412087}.

\bibitem{Lacock:1995tq}
UKQCD, P.~Lacock and C.~Michael,
Phys. Rev. {\bf D52}, 5213 (1995), {\tt hep-lat/9506009}.

\bibitem{Leinweber:2001ac}
D.~B. Leinweber, A.~W. Thomas, K.~Tsushima, and S.~V. Wright,
{\tt hep-lat/0104013}.

\bibitem{Allton:1997yv}
C.~R. Allton, V.~Gimenez, L.~Giusti, and F.~Rapuano,
Nucl. Phys. {\bf B489}, 427 (1997), {\tt hep-lat/9611021}.

\bibitem{Allton:1996kr}
C.~R. Allton, {\tt hep-lat/9610016}.

\bibitem{Allton:1997dn}
C.~R. Allton, Nucl. Phys. Proc. Suppl. {\bf 53}, 867 (1997), {\tt hep-lat/9610014}.

\bibitem{Ukawa:1993hz}
A.~Ukawa, Nucl. Phys. Proc. Suppl. {\bf 30}, 3 (1993).

\bibitem{Ono:1978}
S.~Ono, Phys. Rev. {\bf D17}, 888 (1978).


\end{thebibliography}
\end{document}